Single photon emission from a plasmonic light source driven by a local field-induced Coulomb blockade

Christopher C. Leon[1,*], Olle Gunnarsson[1,*], Dimas G. de Oteyza[2,3,4], Anna Rosławska[1,5], Pablo Merino[1,6,7], Abhishek Grewal[1], Klaus Kuhnke[1,*], Klaus Kern[1,8]

[1]Max-Planck-Institut für Festkörperforschung, Stuttgart, Germany.

[2]Donostia International Physics Center, San Sebastian, Spain.

[3]Centro de Fisica de Materiales, CSIC-UPV/EHU, San Sebastian, Spain.

[4]Ikerbasque, Basque Foundation for Science, Bilbao, Spain.

[5]present address: Université de Strasbourg, CNRS, IPCMS, UMR 7504, F-67000 Strasbourg, France.

[6]Instituto de Ciencia de Materiales de Madrid, CSIC, Madrid, Spain.

[7] Instituto de Física Fundamental, CSIC, Madrid, Spain.

[8]Institut de Physique, École Polytechnique Fédérale de Lausanne, Switzerland.

[*]Corresponding Authors: c.leon@fkf.mpg.de, o.gunnarsson@fkf.mpg.de, k.kuhnke@fkf.mpg.de

## Keywords

## Abstract

A hallmark of quantum control is the ability to manipulate quantum emission at the nanoscale. Through scanning tunneling microscopy induced luminescence (STML) we are able to generate plasmonic light originating from inelastic tunneling processes that occur in a few-nanometer thick molecular film of $C_{60}$ deposited on Ag(111). Single photon emission, not of excitonic origin, occurs with a 1/*e* lifetime of a tenth of a nanosecond or less, as shown through Hanbury Brown and Twiss photon intensity interferometry. We have performed tight-binding calculations of the electronic structure for the combined Ag-$C_{60}$-tip system and obtained good agreement with experiment. The tunneling happens through electric field induced split-off states below the $C_{60}$ LUMO band, which leads to a Coulomb blockade effect and single photon emission. The use of split-off states is shown to be a general technique that has special relevance for narrowband materials with a large bandgap.

## Introduction

Discrete electronic states split-off from Bloch bands have been of interest for decades[1], and continue to be at the forefront of solid state research and development because of their intimate connection with point defects in materials, such as dopant atoms and color centers, which exhibit non-classical effects such as single photon emission. Their optoelectronic properties are well-described by solving the Schrödinger equation for a periodic potential with a local perturbation, with the perturbation depending on the material composition of the defect. Thus, exploring different ways to create and manipulate split-off states are necessary in order to gain tunability beyond this materiality constraint[2,3,4]. Efforts in this direction include quantum dot nanostructures[5,6], but their optoelectronic properties depend on single exciton (electron-hole pair) decay[7] that is not easily modifiable[8].

Thus, a concrete demonstration that single photon emission can be achieved without relying on internal molecular or quantum dot transitions is in order. To this end, we observe that the very origin of split-off states, the local perturbation, need not be materials-based, and in fact, can also be an external perturbation. This motivates using a nanometer tip of a scanning tunneling microscope (STM) as a local perturbation, in a system that can spotlight the principle of single photon emission in a more materials-independent manner.

As a proxy for a generic semiconductor on a metallic substrate, we use a thin $C_{60}$ slab that couples weakly to a metal substrate. Using $C_{60}$ also has the advantage that its electronic structure is accurately represented by a simple model[9], which enables us to rationalize the agreement we obtained between theory and experiment. The narrowness of the $C_{60}$ bands enables split-off states to be created[10] with the electric field of a tip by applying a sufficient bias voltage[11]. In analogy to having a point defect at the tip position, the split-off states tend to be approximately localized on the $C_{60}$ molecule directly beneath the tip, leading to a strong Coulomb repulsion between two electrons in split-off states. This leads to a Coulomb blockade of tunneling events whereby electrons are forced to tunnel one at a time. Consequently, the photon emission via tip-induced plasmons also occurs one at a time[12] whenever it results from tunneling via such states[13].

This phenomenon provides great flexibility and control over single photon emission because it involves electrons tunneling inelastically through a field induced split-off state. Because of the functional dependence of tunneling on tip-sample distance, the intensity or time scale of photon emission can be modified with exponential sensitivity. Moreover, whether or not split-off states are formed below the Fermi energy of the sample, $E_F^S$, is directly controllable by the bias voltage, and is manifestly shown in our measurements. The recovery time of the Coulomb blockade is expected to be strongly influenced by the thickness of the molecular film[14].

None of these methods of control or variability are available when photon emission is due to exciton decay rather than inelastic tunneling. Thus, the performance of this tip-induced quantum dot and single photon emitter in terms of recovery time, without any optimization, is already competitive with epitaxial quantum dot emitters[15]. This work highlights the important role of local electric fields in probe microscopies[16], expands the variety of quantum states that can be made with $C_{60}$[17], and indeed demonstrates the general significance of field-induced split-off states.

## Results

The experiments are performed with an STM with optical access, shown schematically in Figure 1. A voltage bias is applied across the junction formed between the tip and the $C_{60}$-Ag(111) sample. The $C_{60}$ is in the form of a few-monolayer thick film[18,19]. The tunnel current is fixed, and the light originating from tunneling processes is characterized. Its intensity correlation function $g^{(2)}(t)$ is measured with a photon intensity interferometer in a Hanbury Brown and Twiss configuration[20] and its optical spectrum is analyzed in a spectrograph.[21]

The electronic structure of the system is studied theoretically using a tight-binding model of the five HOMO ($h_u$) and three LUMO ($t_{1u}$) states on each $C_{60}$ molecule (see Supplementary Information). The molecules are arranged in laterally infinite layers as a (111) surface, taking into account orientational ordering with four different orientations.[22] The potential from the tip is described by introducing image charges, satisfying the boundary conditions at the tip and the Ag(111)-$C_{60}$ and $C_{60}$-vacuum interfaces. The electronic structure is calculated in a Green's function formalism, assuming that the tip potential affects only a few hundred molecules (see Supplementary Information).

Experimentally, the emitted light spectrum is broad (Figure 2a) and is a signature of plasmonic emission from inelastic scattering events in the tunnel junction.[23] $g^{(2)}(t)$ has a large correlation dip at $t$ = 0 (Figure 2b) and satisfies $g^{(2)}(0) < g^{(2)}(t)$ for all $t > 0$, unlike typical plasmonic emission where $g^{(2)}(t)$ is unity. Hence, the photon emission is antibunched. Based on a 2-particle Hilbert space argument, its dip intensity of 58% (*i.e.* $g^{(2)}(0) = 0.42$) is consistent with imperfect single photon emission of at most $[1-(1-2g^{(2)})^{1/2}]/g^{(2)} = 1.42$ photons on average[24]. The antibunching lifetime is $\tau = 132$ ps and reflects the tunneling of electrons from Ag(111) to the top $C_{60}$ layer. To our knowledge, this is the first demonstration of antibunched plasmonic emission attained in an STM by circumventing the limitations set by emitter lifetimes, whether exciton-based[25,26,27] or otherwise[28].

The photon statistics can be understood by analyzing the behavior of the electronic states. Figure 3a shows the topography of a $C_{60}$ island. The cyan contours mark the terrace boundaries, with the labels indicating the terrace height in $C_{60}$ layers. Figure 3b shows the integrated photon emission efficiency (light emission normalized by current) as a function of tip position. Figure 3c shows the relative tip position and photon emission efficiency along the magenta line in Figures 3a-b. The light emission decreases with layer thickness, then recovers slightly and remains constant at and beyond a critical threshold of four $C_{60}$ layers. Differential conductance (dI/dV) spectra shown in Figure 3d measured from the 1st to 5th terrace show a very pronounced resonance, present at the 4th and 5th terrace for this -3.5 V bias, 20 pA setpoint. The resonance is particularly sharp, suggestive of tunneling through a single electronic state[29]. It is not to be confused with any charge hysteresis in the system[30]. It is assigned to a LUMO-derived state which has been pulled out of the continuum and below $E_F^S$ by the electric field of the tip. We emphasize that this LUMO-derived state appears in experiment at negative rather than positive applied potentials[31].

The controllability of the onset of light emission in the various $C_{60}$ layers is of particular interest. Figure 4a shows the topography of one series of terraces from Figure 3. The thickness threshold at which luminescence occurs is lowered from the 4th to the 2nd layer by changing the bias from -2.86 V to -4 V as seen in Figures 4b and 4c. This effect is examined further by measuring tunneling spectra on the 4th layer as a function of tip-sample distance. Figure 4d shows two resonances later assigned to split-off states from the LUMO band. As the tip-sample distance is increased, the bias has to be made more negative to observe the resonances[32] (Figure 4e).

## Comparison with Calculations

Consider four layers of $C_{60}$ on an ideal metal surface substrate and the projected density of states (DOS) at its topmost layer (Figure 5), with particular focus on the LUMO DOS. It has band edges at 0.27 and 1.02 eV, with first moment 0.73 eV and is concentrated between 0.6 and 1.0 eV. Due to the strong image potential in the innermost layer, the DOS in this layer extends down to 0.27 eV. Orbital mixing between the innermost layer and the topmost layer then leads to the DOS of the topmost layer to have a small tail that extends down to 0.27 eV as well. There is band bending due to the laterally periodic image potential in the substrate, but not due to the localized tip-induced potential. When a small bias is applied with a tip (-0.68 V in Figure 5b), the band edges stay fixed while the local DOS on the $C_{60}$ directly below the tip is depleted at the upper band edge and accumulates at the lower edge. This causes the first moment to move to a lower potential, in this case, towards $E_F^S$ (Figure 5a-b). This is due to the combined effects of the electrostatic and image potentials from the tip. (Details on the HOMO states in Figure 5a-b are available in Supplementary Information.)

For a critical applied voltage $U$, discrete split-off states appear below the LUMO band edge. This critical $U$ exists because any state in a band is at most half a bandwidth in energy away from one band edge. Thus, a split-off state appears when the center of gravity of the DOS on the $C_{60}$ below the tip has moved slightly more than half the $C_{60}$ LUMO bandwidth. This effect is similar to split-off states caused by point impurities in semiconductors, and should not be confused with "band bending" in extended two-dimensional layer systems. Here, the DOS is locally perturbed to the extent that it has little resemblance to the unperturbed DOS.

The split-off states in Figure 5c are localized on the $C_{60}$ molecule below the tip with some extension to the neighboring molecules. The figure shows how the first of a series of LUMO split-off states is formed just below $E_F^S$ for $U$ = -3.33 V and 4 layers, and allows tunneling to the tip through these states. The tunneling contribution from the first split-off state explains the sharp resonance seen in Figure 3d for 4 and 5 ML $C_{60}$. Figure 4d shows the experimental behavior of the first and second split-off states in detail, which correspond to the calculated split-off states with most of their weight in the three LUMO states on the $C_{60}$ molecule below the tip. Of the two split-off states, calculations reveal that the first has more weight because it is doubly degenerate. Note that HOMO split-off states are also present in Figure 5c. These split-off states, and those of Figure 5d, are discussed later.

Now we examine how the LUMO split-off state feature shifts as a function of tunneling parameters and layer thickness. Whenever an increased perturbation occurs across the $C_{60}$ film, extant split-off states will separate further from their parent continuum band. Figure 6a and Figure 6b shows the electrostatic potential for the case $U$ = -3.33 V, fixed sample[33] and tip[34] work function, with varying $C_{60}$ film thickness and vacuum gap size. While the high relative permittivity of $C_{60}$ ($\varepsilon_{bulk}$ = 4.4[35]) causes a large part of the potential drop to occur over vacuum, there is still a substantial drop over the $C_{60}$ film, whose magnitude increases with layer thickness (orange, green, red) and closer tip placement (blue). These behaviors also hold for positive $U$. Thus, in going from two to three layers, the set of split-off states below the HOMO band shift to lower potential, while those above the LUMO band shift to higher potential. Detailed calculations (see Figure 3d and Supplementary Information) reveal the first two split-off states at energies which are in rather good agreement with the applied biases in experiment. However, they also underestimate the energy splitting on the LUMO side possibly due to our neglect of the local crystal field that lowers the symmetry of the $C_{60}$ orbitals at the surface.

Figure 6c shows how the calculated lowest split-off state shifts with applied bias. The case of 4 ML $C_{60}$ that is examined in detail in Figure 6d near $E_F^S$, in particular, shows how the two lowest split-off states depend on $U$ and $d$. The lowest state moves below $E_F^S$ for $U$ = - 3.32 V which is comparable to $U$ = -3.26 V as in experiments. $U$ has to be lowered by an additional 0.22 V to move the second lowest peak through $E_F^S$, which is also comparable to experiment (see Figure 4d). Keeping the peak positions fixed as $d$ is reduced from 0.4 to 0.35 nm requires a shift of 0.17 and 0.19 V, respectively. This gives the derivatives -3.5 (-3.8) V/nm compared with the experimental -1.95 (-2.16) V/nm for the lowest (second lowest) peak and qualitatively reproduces the expected electric field dependence seen in Figure 4d. The approximate factor of 2 discrepancy is considered reasonable given the approximations used in the calculations. Further refinement may require approximations that move beyond a spherical tip geometry, and account for the orbital mixing between the Ag(111) substrate and $C_{60}$. Because the theoretical model of the tip is sharper than a real tip, the calculations are expected to be biased in the direction of having a stronger electric field dependence.[36]

## Discussion

We now discuss the Coulomb blockade that occurs when tunneling happens through a LUMO split-off state, and its relationship with the nonzero photon correlation dip shown in Figure 2b. Consider the DOS on the $C_{60}$ molecule below the tip for $U$ = -3.33 V (Figure 5c, red curve), and the change when a LUMO split-off state is occupied by one electron (Figure 5d, blue curve). Since the DOS are mostly localized on the $C_{60}$ molecule below the tip, the associated Coulomb integrals are large. Thus, the electron occupancy just mentioned increases the electrostatic potential on the $C_{60}$ below the tip by 1.02 eV and shifts the main structure in the LUMO DOS by 1.14 eV. The electrostatic potential on the neighbors of the molecule is increased by 0.34 eV and the DOS is shifted correspondingly to higher energies. These changes prohibit the occupancy of any LUMO state by a second electron (Figure 5d, blue curve). Although this implies that a perfect Coulomb blockade should occur, the photon correlation dip, $g^{(2)}(t = 0)$, is clearly above 0.

To account for $g^{(2)}(t = 0) > 0$, consider concomitant tunneling from the HOMO states. The change between Figure 5c (red curve) and Figure 5d (blue curve) show that the split-off states below the HOMO band edge shift up in energy and become resonances close to the top of the HOMO band at -1.74 eV, when a LUMO state is occupied. This has two consequences. The tunneling rate for a HOMO state is now substantially higher, since the electron has a higher energy when tunneling through the vacuum barrier. Thus, counterintuitively, the Coulomb blockade *helps* the tunneling of a HOMO electron. In addition, this electron may now have enough energy to emit a detectable photon when it decays to the Fermi level of the tip, $E_F^T$, at -3.33 V. If a photon detected at $t = 0$ was emitted by an electron tunneling from the HOMO resonance, a second electron can instantly tunnel from a LUMO split-off state and would reduce the $g^2(t)$ dip at $t = 0$. The resonance in Figure 3d for four layers and $U$ = -3.2 V is then both due to tunneling through the LUMO split-off state *and* an increased tunneling through the HOMO split-off state. We envisage this contribution to be characterized explicitly by performing photon correlation experiments with spectral discrimination filters.

The Coulomb blockade can be overcome by the applied bias when it is lowered beyond $U$ = -6.2 V. Only then can the tip potential move a LUMO split-off state, mainly localized in the innermost $C_{60}$ layer, below $E_F^S$, even if an electron occupies a LUMO split-off state on the $C_{60}$ below the tip. It would be interesting to characterize the behavior of the photon correlation in this situation.

Higher order tunneling mechanisms that occur should also reduce the nominal photon correlation dip. A HOMO electron, from a HOMO split-off state beneath the tip for instance, can tunnel even if the LUMO split-off states are empty. This creates a localized hole whose attractive potential causes the LUMO split-off states beneath the tip and the neighboring molecules to shift below $E_F^S$. The hole mitigates the Coulomb repulsion of housing two electrons within the 6 and 36 states (counting spin and nominal $T_{1u}$ symmetry of the $C_{60}$ LUMO) suddenly made energetically favorable for tunneling. The most important of these trion-like electronic configurations involve an electron on the same molecule as the hole, and an additional electron on the same or neighboring molecule. The latter trion can only be formed by an electron hopping into the neighboring molecule first. The two electrons tunneling once the trion decays would then decrease the correlation dip.

Another possible reason for $g^{(2)}(t = 0) > 0$ is that one tunneling electron can emit two photons[37], which is found to be important, in particular, for large tip-sample distances. We expect this effect to be small here, both because the tip-sample distance is rather small and because the bias is not much larger than twice the lower limit of the detector sensitivity. If two photons are emitted, it is then rather unlikely that both are detected, since energy conservation then requires that both have approximately the energy $|U|/2$.

To highlight some design heuristics for developing plasmonic antibunching emitters beyond $C_{60}$, we have varied the parameters in our model calculations. First, we examine the energetics of making split-off states. We first double the LUMO and HOMO bandwidths, keeping their respective band edges closest to $E_F^S$, fixed relative to $E_F^S$. To align a LUMO split-off state below $E_F^S$ now requires roughly twice as large a bias as before. This proportional relationship favors the use of narrow bandwidth materials.

Next, to favor antibunching, the tunneling contribution from the HOMO states should be lowered, for example, by reducing the large HOMO degeneracy (from 10 in the case of $C_{60}$). Increasing the band gap of $C_{60}$ from 2.3 eV[38] by lowering the HOMO is also favorable. This change would lower the energy of the split-off state above the top of the HOMO that is present when a LUMO split-off state is occupied. This would then reduce the HOMO tunneling to the tip because the electron would now tunnel through a larger barrier, and thus, favor antibunching.

The effective on-site Coulomb interaction, $U_{eff}$, is also important[39], and tends to be smaller for large molecules. Using molecules with a reduced $U_{eff}$ would lower the energy of the HOMO split-off state which again favors antibunching. A material with a larger dielectric function would also reduce $U_{eff}$ through screening effects. However, this change would require a larger bias voltage to make split-off states since the voltage drop over the molecular layers would be reduced. Electronegative molecules, such as $C_{60}$, are favorable in this regard because a smaller bias is needed to pull a state below $E_F^S$.

## Conclusions

We have shown that a local field induced quantum dot can be designed using a molecular film on a metallic substrate in an STM setup. This leads to a Coulomb blockade for tunneling charges and plasmonic photon antibunching. Here a $C_{60}$-Ag(111) system was used, but we also discussed how other molecular characteristics could influence the emitter performance. This approach offers great flexibility, and in particular, an external bias can now regulate whether or not a Coulomb blockade occurs. In conventional quantum dots, the rate of photon emission is determined by the exciton lifetime, which is often hard to influence and much longer than 1 ns. Here, the explicit use of tunneling processes means that the rate has a sensitive exponential dependence on the tip-sample distance, and time scales shorter than 1 ns can be achieved if desired. The recovery time should also depend exponentially on the molecular layer thickness, which can be controllably varied with conventional epitaxial techniques. In summary, the significance of bypassing the use of excitons in single photon emission is that the rate of single photon production, or equivalently, the intensity, is no longer tied to the exciton lifetime. We anticipate future work to develop in these directions.

## Methods

To make the samples, $C_{60}$ is vapor deposited onto Ag(111) from a Knudsen cell for 1 hour, with the sample temperature set between 213-253 K. This temperature range favors rough island growth, which then facilitates measuring the light emission as a function of layer thickness and STM parameters within a small rastered area. The resulting $C_{60}$-Ag(111) topography and its characterization is shown in panels a-d of Figure 3. The Ag(111) surface itself was prepared with standard cycles of $Ar^+$ sputtering at 300 K followed by annealing at 900 K.

## Acknowledgements

D. G. O. acknowledges support from the Alexander von Humboldt Foundation for his research stay at the MPI, hosted by K. Ke., as well as from the European Research Council under the European Union's Horizon 2020 research and innovation programme (grant agreement No. 635919) and from the Spanish Ministry of Economy, Industry and Competitiveness (MINECO, Grant No. MAT2016-78293-C6-1-R).

## Author Contributions

C. C. L., D. G. O., A. R., P. M., A. G., and K. Ku. performed the experimental work. C. C. L. analyzed the experimental data. O. G. performed the tight-binding calculations for the electronic structure of the Ag-$C_{60}$-tip system. C. C. L. and O. G. wrote the manuscript with input from all authors. K. Ku. and K. Ke. conceived and designed the research program.

## Competing Interests

The authors declare that they have no competing interests.

## Materials & Correspondence

Correspondence to C. C. L., O. G., or K. Ku.

# References

[1] Slater, J. C. *The Electronic Structure of Solids*. In *Electrical Conductivity I / Elektrische Leitungsphänomene I. Series: Encyclopedia of Physics / Handbuch der Physik.* Editor Flügge, S., vol. 19, 1-136. (Springer, Berlin, 1956).

[2] Yuan, Z. et al. Electrically driven single-photon source. *Science* **4**, 102-105 (2002).

[3] He, Y.-M. et al. Single quantum emitters in monolayer semiconductors. *Nat. Nanotech.* **10**, 497-502 (2015).

[4] Chakraborty, C., Kinnischtzke, L., Goodfellow, K. M., Beams, R., Vamivakas, A. N. Voltage-controlled quantum light from an atomically thin semiconductor. *Nat. Nanotech.* **10**, 507-511 (2015).

[5] Coe, S., Woo, W.-K., Bawendi, M., Bulović, V. Electroluminescence from single monolayers of nanocrystals in molecular organic devices. *Nature* **420**, 800-803 (2002).

[6] Kroupa, D. M. et al. Tuning colloidal quantum dot band edge positions through solution-phase surface chemistry modification, *Nat. Commun.* **8**, 15257 (2017).

[7] Aharonovich, I., Englund, D., Toth, M. Solid-state single-photon emitters. *Nat. Photonics* **10**, 631-641 (2016).

[8] Tonndorf, P. et al. Single-photon emission from localized excitons in an atomically thin semiconductor. *Optica* **2**, 347-352 (2015).

[9] Gunnarsson, O., Erwin, S. C., Koch, E., Martin, R. M. Role of alkali atoms in $A_4C_{60}$. *Phys. Rev B* **57**, 2159-2162 (1998).

[10] Blossey, D. F. Wannier exciton in an electric field. I. Optical absorption by bound and continuum states. *Phys. Rev. B* **2**, 3976-3990 (1970).

[11] Dombrowski, R., Steinebach, Chr., Wittneven, Chr., Morgenstern, M., Wiesendanger, R. Tip-induced band bending by scanning tunneling spectroscopy of the states of the tip-induced quantum dot on InAs(110). *Phys. Rev. B* **59**, 8043-8048 (1999).

[12] Imamoğlu A., Yamamoto, Y. Nonclassical light generation by Coulomb blockade of resonant tunneling. *Phys. Rev. B* **46**, 15982-15992 (1992).

[13] Bauer, G. E. W. Excitonic correction to resonant tunneling. *Surf. Sci.* **305**, 358-362 (1994).

[14] Rosławska, A. et al. Single charge and exciton dynamics probed by molecular-scale-induced electroluminescence. *Nano Lett.* **18**, 4001-4007 (2018).

[15] Senellart, P., Solomon, G., White, A. High-performance semiconductor quantum-dot single-photon sources. *Nat. Nanotechnol.* **12**, 1026-1039 (2017).

[16] Wagner, C. et al. Scanning quantum dot microscopy. *Phys. Rev. Lett.* **115**, 026101 (2015).

[17] Zhu, X.-Y. et al. Molecular quantum well at the $C_{60}$/Au(111) interface. *Phys. Rev. B* **74**, 241401(R) (2006).

[18] Große, C. et al. Submolecular electroluminescence mapping of organic semiconductors. *ACS Nano* **11**, 1230-1237 (2017).

[19] Merino, P., Große, C., Rosławska, A., Kuhnke, K., Kern, K. Exciton dynamics of $C_{60}$-based single-photon emitters explored by Hanbury Brown-Twiss scanning tunnelling microscopy. *Nat. Commun.* **6**, 8461 (2015).

[20] Hanbury Brown, R. & Twiss, R. Q. Interferometry of the intensity fluctuations in light II. An experimental test of the theory for partially coherent light. *Proc. R. Soc. London, Ser. A* **243**, 291-319 (1958).

[21] Kuhnke, K. et al. Versatile optical access to the tunnel gap in a low-temperature scanning tunneling microscope. *Rev. Sci. Instrum.* **81**, 113102 (2010).

[22] David, W.I.F. et al. Crystal structure and bonding of ordered $C_{60}$. *Nature* **353**, 147-149 (1991).

[23] Merino, P. et al. Bimodal exciton-plasmon light sources controlled by local charge carrier injection. *Sci. Adv.* **4**, eaap8349 (2018).

[24] Zubizarreta Casalengua, E., López Carreño, J. C., del Valle, E., Laussy, F. P. Structure of the harmonic oscillator in the space of n-particle Glauber correlators. *J. Math. Phys.* **58**, 062109 (2017).

[25] Akimov, A. V. et al. Generation of single optical plasmons in metallic nanowires coupled to quantum dots. *Nature* **450**, 402-406 (2007).

[26] Zhang, L. et al. Electrically driven single-photon emission from an isolated single molecule. *Nat. Commun.* **8**, 580 (2017).

[27] Luo, Y. et al. Electrically driven single-photon superradiance from molecular chains in a plasmonic nanocavity. *Phys. Rev. Lett.* **122**, 233901 (2019).

[28] Kolesov, R. et al. Wave-particle duality of single surface plasmon polaritons. *Nat. Phys.* **5**, 470-474 (2009).

[29] Kalmeyer, V., Laughlin, R. B. Differential conductance in three-dimensional resonant tunneling. *Phys. Rev. B* **35**, 9805-9808 (1987).

[30] Buot, F. A. Emitter quantization and double hysteresis in resonant-tunneling structures: a nonlinear model of charge oscillation and current instability. *Phys. Rev. B* **61**, 5644-5665 (2000).

[31] Nazin, G. V., Wu, S. W., Ho, W. Tunneling rates in electron transport through double-barrier molecular junctions in a scanning tunneling microscope. *Proc. Nat. Acad. Sci.* **102**, 8832-8837 (2005).

[32] Große, C. et al. Dynamic control of plasmon generation by an individual quantum system. *Nano Lett.* **14**, 5693-5697 (2014).

[33] Veenstra, S. C., Heeres, A., Hadziioannou, G., Sawatzky, G. A., Jonkman, H. T. On interface dipole layers between $C_{60}$ and Ag or Au. *Appl. Phys. A* **75**, 661-666 (2002).

[34] Hansson, G. V. & Flodström, S. A. Photoemission study of the bulk and surface electronic structure of single crystals of gold. *Phys. Rev. B* **18**, 1572-1585 (1978).

[35] Hebard, A. F., Haddon, R. C., Flemming, R. M., Kortan, R. Deposition and characterization of fullerene films. *Appl. Phys. Lett.* **59**, 2109-2111 (1991).

[36] Urbieta, M. et al. Atomistic-scale lighting rod effect in plasmonic picocavities: a classical view to a quantum effect. *ACS Nano* **12**, 585-595 (2018).

[37] Leon, C. C. et al. Photon superbunching from a generic tunnel junction. *Sci. Adv.* **5**, eaav4986 (2019).

[38] Lof, R. W., van Veenendaal, M. A., Koopmans, B., Jonkman, H. T., Sawatzky, G. A. Band gap, excitons, and Coulomb interaction in solid $C_{60}$. *Phys. Rev. Lett.* **68**, 3924-3927 (1992).

[39] Antropov, V. P., Gunnarsson, O., Liechtenstein, A. I. Phonons, electron-phonon, and electron-plasmon coupling in $C_{60}$ compounds. *Phys. Rev. B* **48**, 7651-7664 (1993).

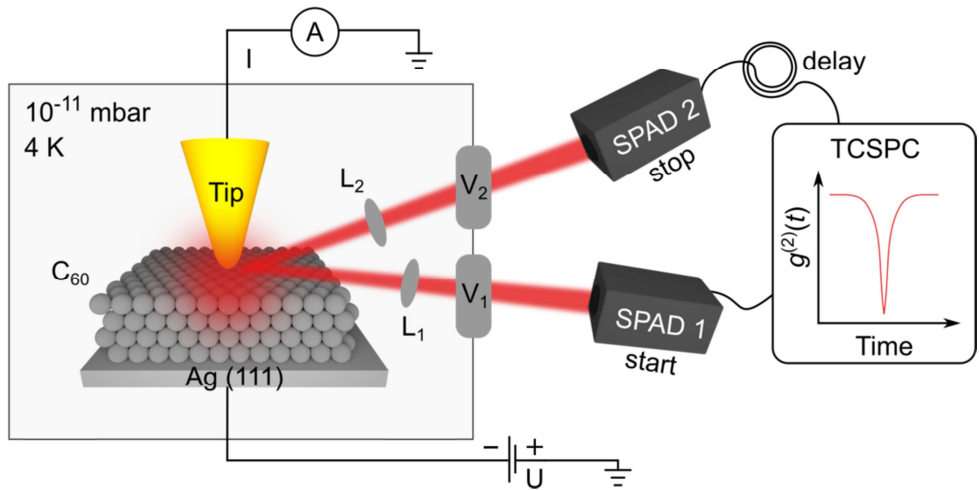

**Figure 1. A scanning tunneling microscope combined with a Hanbury Brown and Twiss interferometer.** Light radiating from a junction formed between a Au tip and multilayer $C_{60}$-Ag(111) substrate travels along optical paths (1, 2) containing (L)enses and (V)iewports to a pair of single-photon avalanche detectors (SPADs). The number of photon coincidence events as a function of time delay $t$ between the SPADs, $g^{(2)}(t)$, is measured with a time-correlated single-photon counter (TCSPC). The voltage bias (U) is applied to the substrate. The tunnel current (I) is measured with a picoammeter (A). A third optical path to an optical spectrometer is not shown.

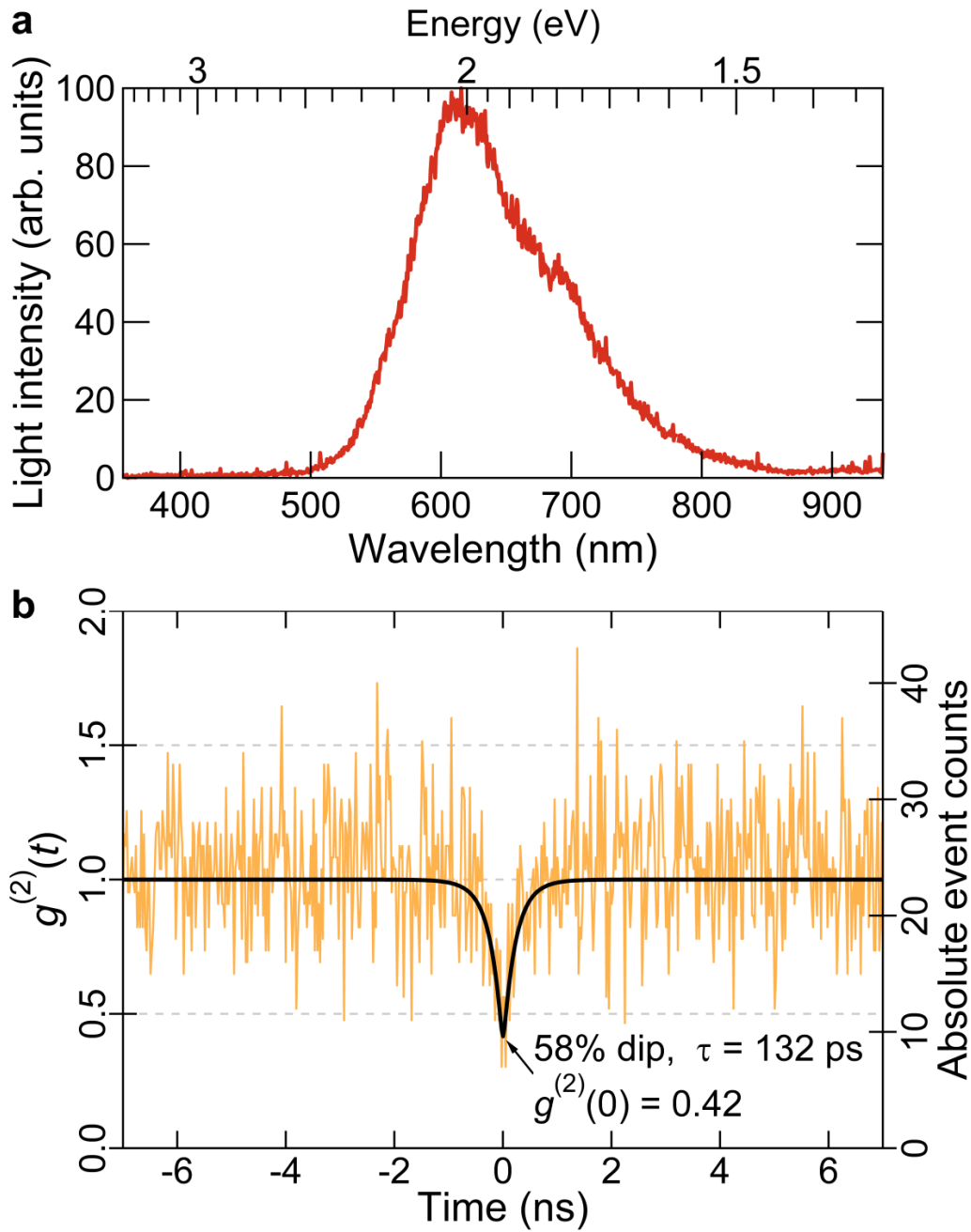

**Figure 2. Plasmonic antibunching light emission from $C_{60}$ multilayers on Ag(111) measured at 125 pA, -3.5 V.** (a) Spectrum measured on a $C_{60}$ terrace, 1200 s. (b) Photon intensity correlation. The level of coincidence events that corresponds to lim $t \rightarrow$ infinity $g^{(2)}(t) = 1$ is proportional to the product of the two SPAD count rates. 7500 s accumulation time, 24.4 ps bin width. The black curve is a least squares Poisson-weighted fit to the data, using a constant background, and the detector response convolved with a Laplace distribution, whose decay parameter τ is defined as the antibunching lifetime.

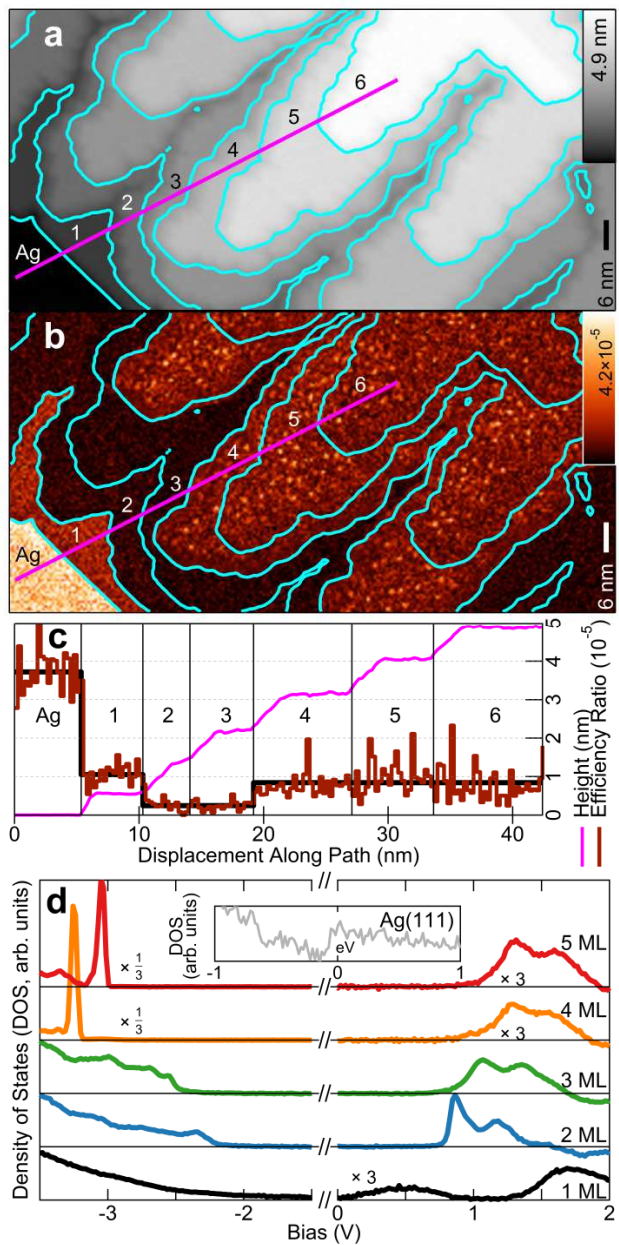

**Figure 3. STML surface characterization of $C_{60}$-Ag(111).** (a) Topography of a $C_{60}$ film on Ag(111); $U$ = -3.5 V, I = 20 pA constant current. (b) Light emission map, normalized to the current, calculated by taking the number of photons per second, and dividing by the number of electrons per second. (c) Line profiles along the magenta lines marked in (a) and (b). The thick black line is a guide to the eye. (d) Differential conductance (dI/dV) spectra of the substrate (inset) and of 1-5 ML $C_{60}$ atop this substrate, with spectra shifted vertically for clarity.

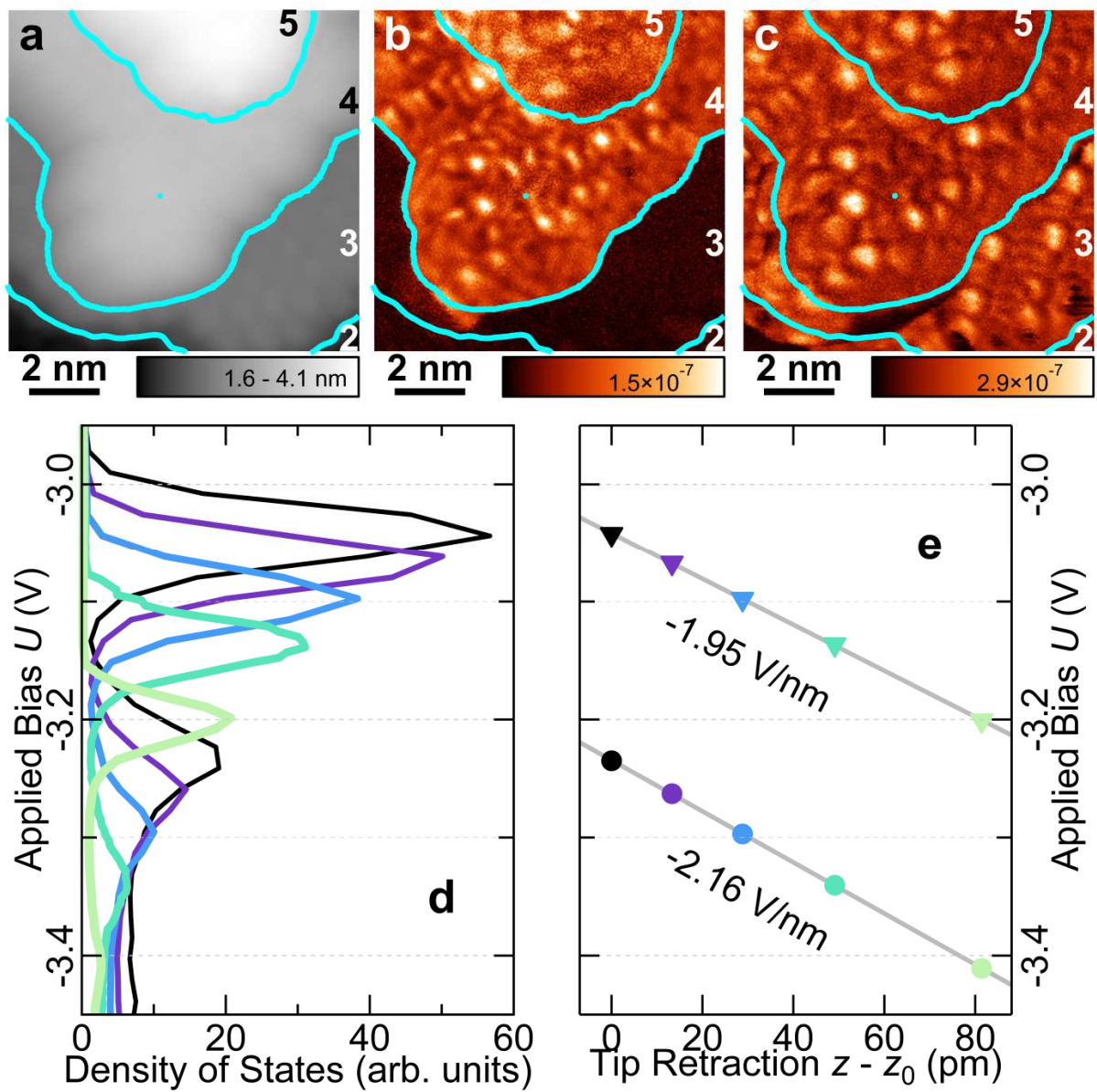

**Figure 4. Controllability of resonant tunneling luminescence**. Measurements of 2-5 monolayers of $C_{60}$ on Ag(111) at 100 pA constant current, with terrace boundary contours marked in cyan. (a, b) Topography and integrated photon emission intensity, both at -2.86 V. Intensity defined as the number of photons per second, divided by the number of electrons per second. (c) Integrated photon emission intensity at -4.00 V. (d) Differential conductance (dI/dV) spectra on the 4th $C_{60}$ terrace as a function of relative tip-sample distance. Two LUMO resonance features are visible. (e) The energetic position of the resonances as a function of relative tip-sample distance. The lines are a fit to the data points and show that the resonance positions follow lines of constant electric field.

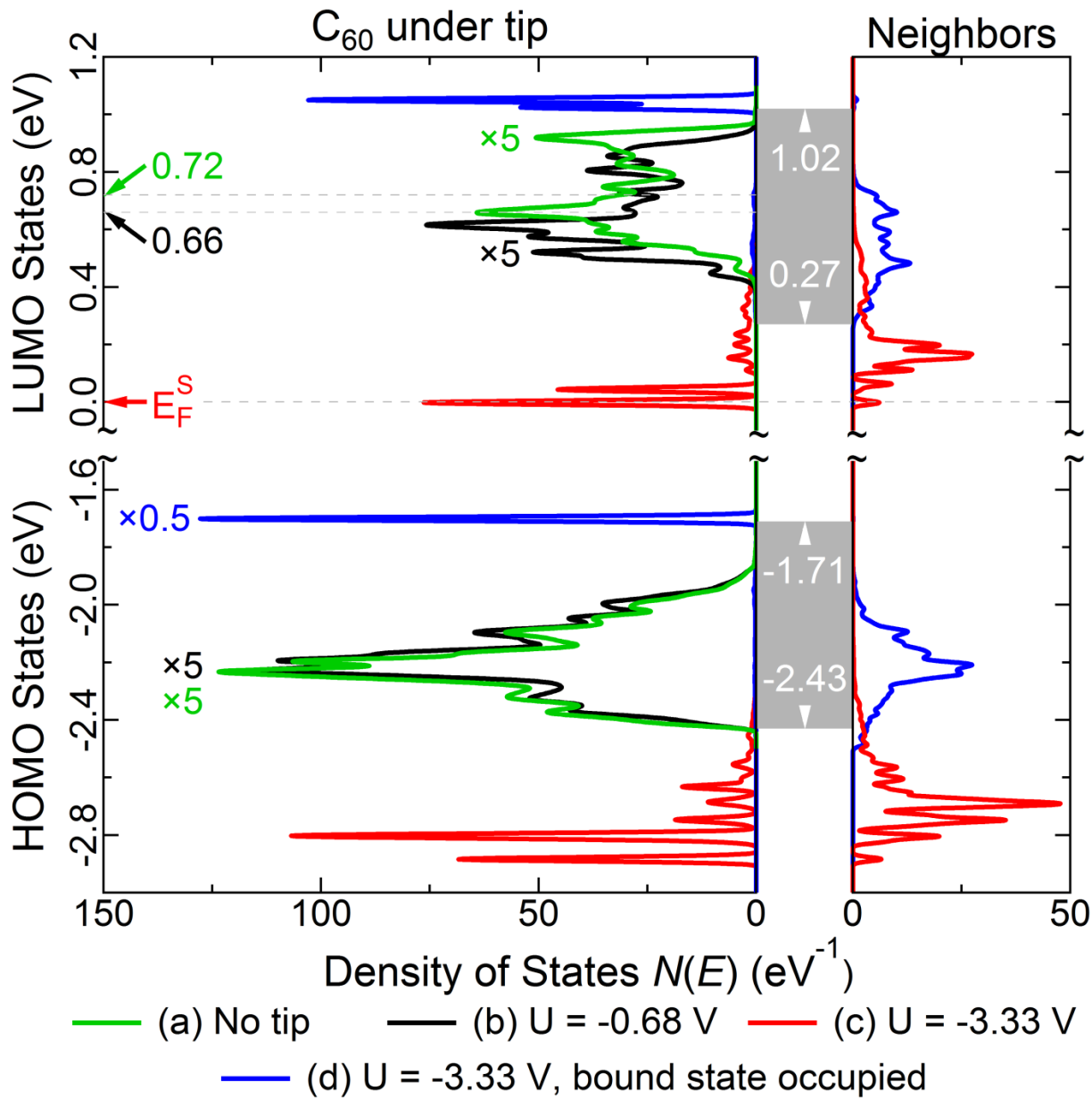

**Figure 5. Calculated projected density of states of 4 ML $C_{60}$-Ag(111) as a function of applied bias.** Grey rectangles indicate the energy range and the onset values of the continuum band of $C_{60}$ states. The band edges stay fixed and do not change with applied bias. Features beyond the band edge correspond to discrete split-off states. Sharp features overlapping with the continuum band are resonances. Arrows along the left vertical axis indicate the band energy in the absence of hybridization. All features are broadened with a FWHM Lorenzian of 0.02 eV. The green curves are projected density of states of the topmost layer.

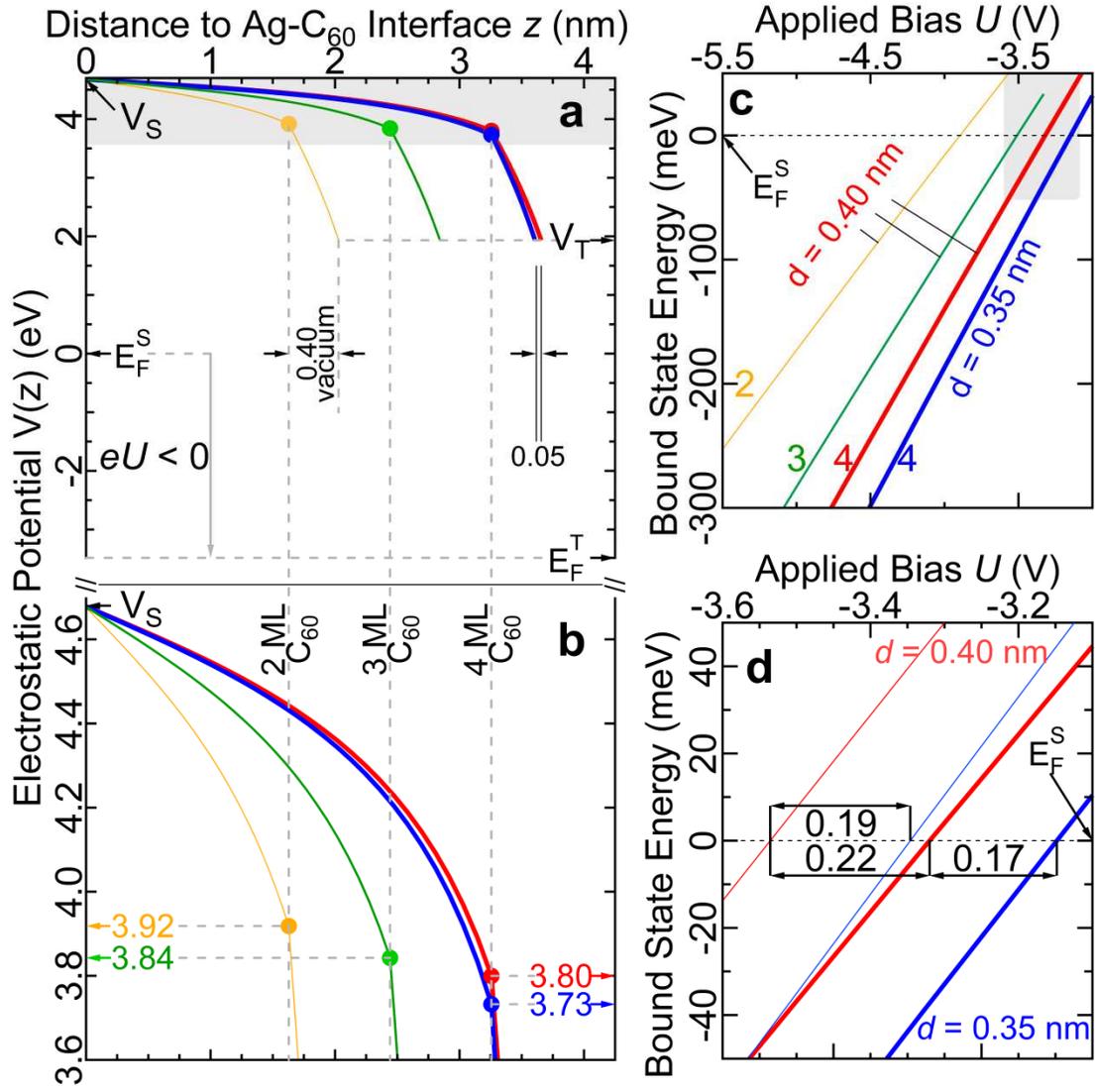

**Figure 6. Calculated electrostatic potential and resultant split-off state potential energies.** The orange, green, red, and blue curves are for a vacuum gap of $d$ = 0.40, 0.40, 0.40, 0.35 nm between the tip and $C_{60}$ surface, and a layer thickness of 2, 3, 4, 4 ML $C_{60}$, respectively. The sample and tip Fermi levels ($E_F^S$ and $E_F^T$), work functions ($\Phi_S$, $\Phi_T$) and vacuum levels ($V_S, V_T$) satisfy $V_S = E_F^S + \Phi_S$ and $V_T = E_F^T + \Phi_T$. a) The potential directly beneath the tip for the bias $U$ = -3.33 V. The Ag(111)-$C_{60}$ interface is at $z = 0$. The solid circles denote the $C_{60}$-vacuum interface. b) Detail of the shaded region in a). c) The lowest LUMO split-off state energy as a function of $U$. d) Detail of the shaded region in c) with additionally, the second lowest singly-degenerate LUMO split-off state shown with thin lines. The double-headed arrows indicate the voltage difference between where the curves cross zero energy, corresponding to $E_F^S$.

Single photon emission from a plasmonic light source driven by a local field-induced Coulomb blockade

Christopher C. Leon[1,*], Olle Gunnarsson[1,*], Dimas G. de Oteyza[2,3,4], Anna Rosławska[1,5], Pablo Merino[1,6,7], Abhishek Grewal[1], Klaus Kuhnke[1,*], Klaus Kern[1,8]

[1]Max-Planck-Institut für Festkörperforschung, Stuttgart, Germany.

[2]Donostia International Physics Center, San Sebastian, Spain.

[3]Centro de Fisica de Materiales, CSIC-UPV/EHU, San Sebastian, Spain.

[4]Ikerbasque, Basque Foundation for Science, Bilbao, Spain.

[5]present address: Université de Strasbourg, CNRS, IPCMS, UMR 7504, F-67000 Strasbourg, France.

[6]Instituto de Ciencia de Materiales de Madrid, CSIC, Madrid, Spain.

[7] Instituto de Física Fundamental, CSIC, Madrid, Spain.

[8]Institut de Physique, École Polytechnique Fédérale de Lausanne, Switzerland.

[*]Corresponding Authors: c.leon@fkf.mpg.de, o.gunnarsson@fkf.mpg.de, k.kuhnke@fkf.mpg.de

# Supplementary Information

## Model

The $C_{60}$ film is described as a few (*n*) layers of $C_{60}$ molecules forming a (111) surface whose molecular positions and orientations are obtained from bulk crystalline $C_{60}$[1]. The film's electronic structure is based on a tight-binding parameterization of 4*n* molecules per unit cell volume that is periodically continued to form an infinite film parallel to the surface. The factor 4 accounts for the four different molecular orientations of $C_{60}$ that are the observed positions found to be frozen in at 4 K, the experimental temperature. One 2*p* orbital is placed on each C atom, pointing radially out from the molecule. The five HOMO ($h_u$) and three LUMO ($t_{1u}$) frontier orbitals of each molecule are then constructed from these 2*p* orbitals. The hopping integrals between the 2*p* orbitals are parameterized by fitting them to DFT calculations[2] from which the hopping integrals between the HOMO and LUMO orbitals on different molecules as well as the potential energies of these orbitals are then calculated. This results in a very accurate representation of the LUMO DFT band structure.[2] The electronic structure calculation of the full film then involves (8 × 4*n*) by (8 x 4*n*) matrices. The factor 8 accounts for the $h_u$ and $t_{1u}$ orbitals of $C_{60}$.

Electrostatic effects are gradually added to the model, starting with the STM tip being absent. In this context the $C_{60}$ layer is treated as a homogeneous dielectric medium with a dielectric constant of ε=4.4[3] that rests on a perfect metal surface. The $C_{60}$ layer thickness is $na/\sqrt{3}$, where *a* is the $C_{60}$ lattice parameter and the factor $1/\sqrt{3}$ is due to the crystal face being (111). The image charges due to an electron in the LUMO or a hole in the HOMO are now considered. An electron in the LUMO on molecule *i* feels an image potential $-U_i^{\text{image}}$, while a hole in the HOMO feels the potential $+U_i^{\text{image}}$. The effect of the vacuum-$C_{60}$ and $C_{60}$-metal interfaces are taken into account using image charges and the appropriate boundary conditions for these interfaces. Image charges are chosen such that the boundary conditions are satisfied for one surface at a time. This approach is iterated until the process converges and the boundary conditions are finally satisfied at both surfaces.

Finally, we take the tip into account. The tip is described as a sphere with radius *R* = 1 nm whose surface is a distance *d* away from the vacuum-$C_{60}$ interface. A charge $Q = U_t R$ is added at the sphere's center to obtain the potential $U_t$ at the sphere's surface. Image charges are then introduced once again to restore the correct boundary conditions at the tip-vacuum, vacuum-$C_{60}$, and $C_{60}$-metal interfaces. The resulting potential at molecule *i* is $U_i^{\text{tip}}$. Finally, we also include the image potential of an electron (in the LUMO) or a hole (in the HOMO) in the tip, obtaining slightly different tip potentials for electrons and holes. $U_i^{\text{image}}$ only depends on the layer index $i$, while $U_i^{\text{tip}}$ also has a strong lateral dependence and is assumed to be localized to an area that would encompass about 100 $C_{60}$ molecules per layer.

The work function for a multilayer $C_{60}$ film on polycrystalline Ag is 4.68 eV[4] and for Au it is 5.26 eV.[5] These values are used to construct and approximate the potential between the $C_{60}$-Ag(111) sample and the Au tip. The difference between these work functions is 0.58 eV. While this difference is sensitive to Ag atoms coating the Au tip during the experimental tip preparation, this effect is not taken into account because the qualitative features of the resulting calculations are relatively insensitive to this difference. Thus, when the bias $U$ is applied, we simply require that $U_t = U + 0.58$ eV.

The tight-binding band gap of $C_{60}$ is 2.24 eV, reasonably close to the experimental result 2.3 eV.[6] We have then applied a small correction to the $C_{60}$ HOMO band, $\Delta E_g$ = - 0.06 eV. This approach puts the bottom of the LUMO band for the four layer system at about 0.27 eV, slightly above $E_F^S = 0$. This alignment to $E_F^S$ is arbitrary, but leads to a LUMO bound state going through $E_F^S$ for approximately correct biases for one to four layer systems.

For a bias of $U < -3.32$ V, a split-off LUMO state moves below $E_F^S$ and can be occupied. If this split-off state has the weight $q_i$ on molecule $i$, we add these charges to the center of the molecule and calculate their images in the surfaces. This leads to an additional potential, $U_i^{\text{filled}}$, resulting from all the charges $q_i$ and their images, except that the potential at molecule $i$ from charge $q_i$ is $q_i U_{\text{eff}}$. Here, $U_{\text{eff}} = 1.3$ V is an effective Coulomb interaction.[7] This calculated value has been shown to be in good agreement with experiment.[6] If the split-off state is assumed to be localized on one molecule in the outermost layers, the net effect of $U_{\text{eff}}$ and the corresponding image charges is an overall shift of 1.5 V, comparable with the experimental value 1.6 V.[6]

The resulting Hamiltonian is then $H = H_0 + U$ with

$$H_0 = \sum_{iv\sigma} \varepsilon_{iv} \hat{n}_{iv\sigma} + \sum_{ivj\mu\sigma} t_{ivj\mu} \left( \hat{c}_{iv\sigma}^\dagger \hat{c}_{j\mu\sigma} + \hat{c}_{j\mu\sigma}^\dagger \hat{c}_{iv\sigma} \right)$$

$$U = \sum_{iv\sigma} U_{iv}^{\text{tip}} \hat{n}_{iv\sigma},$$

Here $\varepsilon_{i\nu} = \varepsilon_{i\nu}^0 + U_i^{\text{image}}$ for $\nu = 1,\cdots,5$ ($\varepsilon_{i\nu} = \varepsilon_{i\nu}^0 - U_i^{\text{image}}$ for $\nu = 6,7,8$) is the position of the HOMO (LUMO) level for molecule $i$. $U_{i\nu}^{\text{tip}}$ has a $\nu$ dependence, since the image potential has different signs for electrons and holes. $\hat{c}_{i\nu\sigma}^\dagger$ and $\hat{c}_{i\nu\sigma}$ are the creation and annihilation operators for an electron in orbital $\nu$ with spin $\sigma$ on molecule $i$, and $\hat{n}_{i\nu\sigma} = \hat{c}_{i\nu\sigma}^\dagger \hat{c}_{i\nu\sigma}$. $t_{i\nu j\mu}$ is the hopping integral between orbital $\nu$ on molecule $i$ and orbital $\mu$ on molecule $j$. If one split-off LUMO state is filled, we also add the potential $U_i^{\text{filled}}$ from the electron in this state.

## Numerical Methods

The electronic structure is calculated using a Green's function technique. First the unperturbed Green's function $G_{i\nu j\nu'}^0(\varepsilon)$ as a function of energy $\varepsilon$ is calculated for a film with an infinite lateral extension, taking the laterally periodic terms $H_0$ (and $U_i^{\text{image}}$) into account. Then we solve the Dyson equation for the perturbed Green's function $G_{i\nu j\nu'}(\varepsilon)$ where

$$G_{i\nu j\nu'}(\varepsilon) = G_{i\nu j\nu'}^0(\varepsilon) + \sum_{k\nu''} G_{i\nu k\nu''}^0(\varepsilon) \cdot U_{k\nu''}^{\text{tip}} \cdot G_{k\nu'' j\nu'}(\varepsilon)$$

We assume that the tip potential is non-negligible over $N_{\text{pert}}$ molecules. Thus, the equations describing $G_{i\nu j\nu'}(\varepsilon)$ involves $(8N_{\text{pert}}) \times (8N_{\text{pert}})$ matrices, where the factor 8 again originates from the frontier orbitals of C$_{60}$. The density of states (DOS), $N_{i\nu}(\varepsilon)$, projected on an orbital $i\nu$ is then

$$N_{i\nu}(\varepsilon) = \frac{1}{\pi} \text{Im}\, G_{i\nu i\nu}(\varepsilon - i0^+)$$

where $0^+$ is an infinitesimal positive quantity. The DOS has contributions both inside the C$_{60}$ HOMO and LUMO bands as well as from split-off states outside of these bands.

## Electronic structure

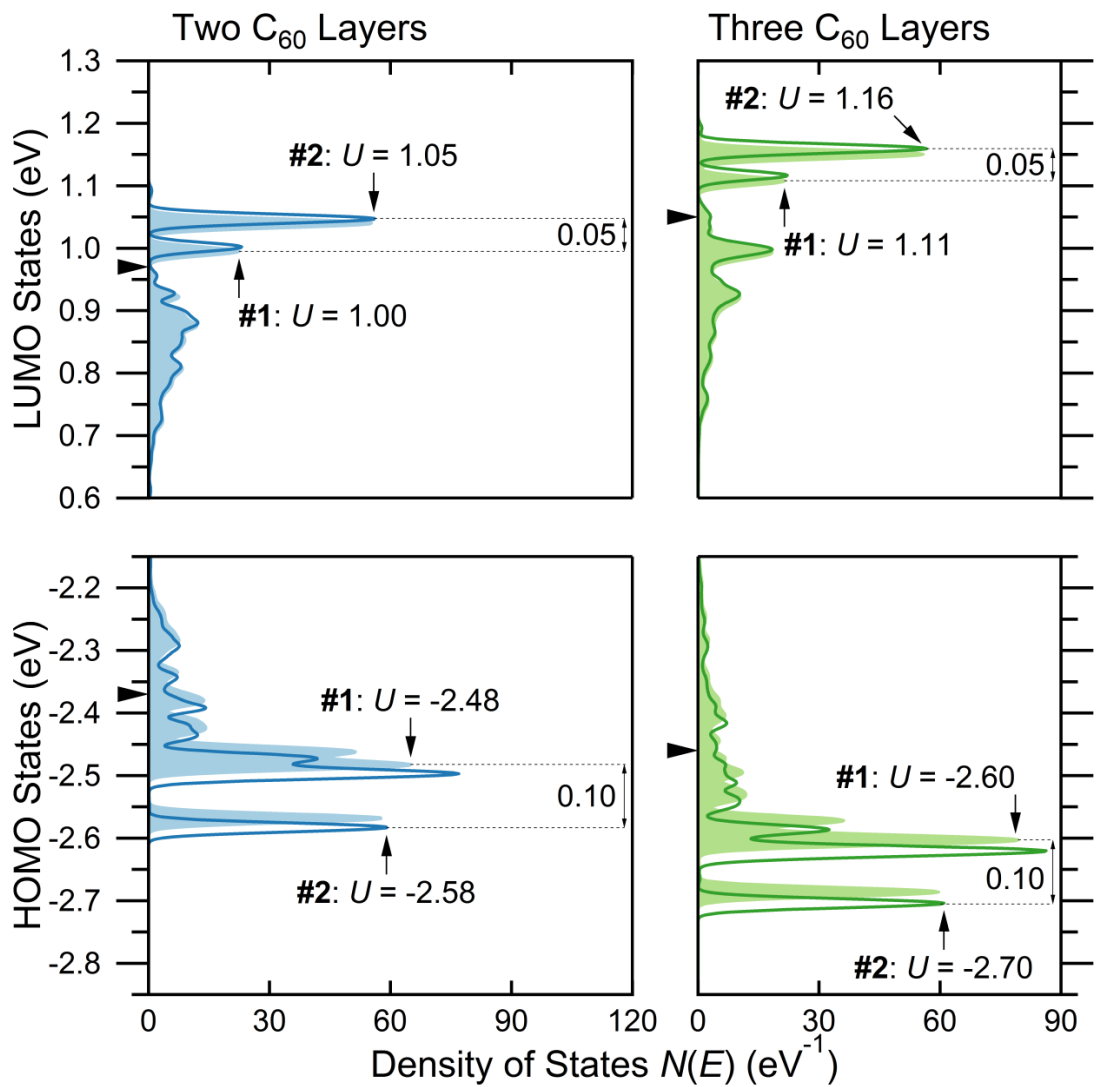

**Supplementary Figure 1. Calculated total density of states for two and three C$_{60}$ layers at the labeled bias values *U* and tip distance *d* = 0.4 nm.** The filled and unfilled curves represent the density of states at the onset bias *U* where tunneling through the first (#1) and second (#2) split-off state is possible, with the specific threshold *U* values indicated beside the arrows. The voltage differences between the two split-off states for each calculated system are indicated with double arrows. All features are broadened with a FWHM Lorentzian of 0.02 eV.

Supplementary Figure 1 shows the electronic structure of the HOMO and LUMO bands for two and three $C_{60}$ layers. These structures were calculated for values of $U$ for which tunneling from a new split-off state or narrow resonance becomes possible. As a specific example, in order to tunnel through the first HOMO split-off state for 3 layers of $C_{60}$, an applied bias of -2.60 V is required. The curve that is labeled with "#1: -2.60" corresponds to the calculated density of states for an applied bias of -2.60 V. These values of $U$ then correspond to structures in the $dI/dU$ curves in Figure 3d. The positions of the structures agree well with experiment, except that the splitting between the peaks in the LUMO (and perhaps the HOMO) is too small compared with experiment (see Figure 3d in main text). This could be due to the neglect of crystal-field splittings in the model, specifically, the splitting of the $t_{1u}$ and $h_u$ states due to the presence of a surface, and the applied potential having a different effect on each orbital because they are oriented differently with respect to the tip and surface. In addition, the tunneling matrix elements may coincidentally cause some features to be obscured during measurements.

## Qualitative behavior of HOMO and LUMO States in Figure 5 at small bias

As stated in the main text: "When a small bias is applied with a tip (-0.68 V in Figure 5b), the band edges stay fixed while the local DOS on the $C_{60}$ directly below the tip is depleted at the upper band edge and accumulates at the lower edge. This causes the first moment to move to a lower potential, in this case, towards $E_F^S$ (Figure 5a-b). This is due to the combined effects of the electrostatic and image potentials from the tip."

For the HOMO these contributions have opposite signs which happen to shift the HOMO slightly upwards.

The absolute energy shift of the density of states depends on both the electrostatic potential and the image potential, and their relative magnitudes. For Ag(111)-$C_{60}$ it is the case that the HOMO DOS shifts positively in energy, in an opposite direction to the small negatively applied bias of -0.68 V.

[1] David, W.I.F. et al. Crystal structure and bonding of ordered $C_{60}$. *Nature* **353**, 147-149 (1991).
[2] Gunnarsson, O., Erwin, S. C., Koch, E., Martin, R. M. Role of alkali atoms in $A_4C_{60}$. *Phys. Rev B* **57**, 2159-2162 (1998).
[3] Hebard, A. F., Haddon, R. C., Flemming, R. M., Kortan, R. Deposition and characterization of fullerene films. *Appl. Phys. Lett.* **59**, 2109-2111 (1991).
[4] Veenstra, S. C., Heeres, A., Hadziioannou, G., Sawatzky, G. A., Jonkman, H. T. On interface dipole layers between $C_{60}$ and Ag or Au. *Appl. Phys. A* **75**, 661-666 (2002).
[5] Hansson, G. V. & Flodström, S. A. Photoemission study of the bulk and surface electronic structure of single crystals of gold. *Phys. Rev. B* **18**, 1572-1585 (1978).

[6] Lof, R. W., van Veenendaal, M. A., Koopmans, B., Jonkman, H. T., Sawatzky, G. A. Band gap, excitons, and Coulomb interaction in solid $C_{60}$. *Phys. Rev. Lett.* **68**, 3924-3927 (1992).

[7] Antropov, V. P., Gunnarsson, O., Liechtenstein, A. I. Phonons, electron-phonon, and electron-plasmon coupling in $C_{60}$ compounds. *Phys. Rev. B* **48**, 7651-7664 (1993).